\documentclass[aps,showpacs]{revtex4}

\usepackage{amssymb}
\usepackage{amsmath}
\usepackage[dvips]{epsfig}

\begin{document}
\title{Electromagnetic wave propagation in an almost circular bundle of closely packed,
metallic, carbon nanotubes}

\author{M. V. Shuba, S. A. Maksimenko}
\affiliation{\\ Institute for Nuclear Problems, Belarus State
University, Bobruiskaya 11, 220050 Minsk, Belarus}
\author{A. Lakhtakia}
\affiliation{\\
Department of Engineering Science and Mechanics,
Pennsylvania State University, University Park, PA 16802-6812,
USA}
\begin{abstract}

An equivalent-multishell approach   for the approximate
calculation of the characteristics of electromagnetic waves
propagating in almost circular (azimuthally symmetric), closely
packed  bundles of parallel, identical, and metallic carbon
nanotubes (CNTs) yields results in reasonably good agreement with
a many-body technique, for infinitely long bundles when the number
of CNTs is moderately high. The slow-wave coefficients for
azimunthally symmetric guided waves increase with the number of
metallic CNTs in the bundle, tending  for thick bundles to unity,
which is characteristic of macroscopic metallic wires. The
existence of an azimuthally nonsymmetric guided wave at low
frequencies in a bundle of a large number of finite-length CNTs
stands in contrast to the characteristics of guided-wave
propagation in a single CNT. The equivalent-multishell approach
yields the polarizability scalar and the antenna efficiency of a
bundle of  finite-length CNTs in the long-wavelength regime  over
a wide frequency range spanning the terahertz and  the
near-infrared regimes. Edge effects give rise to geometric
resonances in such bundles.
 The
antenna efficiency  of a CNT bundle at the first resonance can
exceed that of a single CNT by four orders
of magnitude, which is promising
 for the design and development of CNT-bundle
antennas and  composite materials   containing CNT-bundles as
inclusions.

\end{abstract}
 \maketitle

\section{Introduction}

Carbon nanotubes (CNTs) have been the subjects of
 intensive research in
 the area of nanotechnology for about 15 years \cite{WL06}, yet
understanding of their fundamental physics   is far from complete.
One topic under intensive investigation is their electromagnetic
response. Early theoretical studies on CNTs, modeled as infinitely
long cylinders of a gas of electrons, showed the existence of
gapless, low-frequency, plasmon branches \cite{Longe}.  The
low-frequency plasmon excitation along the nanotube axis lead to
the formation of a electromagnetic surface wave \cite{Slepyan99}.
This surface wave strongly influences the scattering
\cite{Slepyan06} and radiation \cite{Hanson05, Burke06} properties
of a CNT in the terahertz regime.

Guided-wave propagation should therefore occur in a CNT bundle
containing 2 $\sim $ 1000 parallel CNTs closely packed together. A guided
wave could be formed by the plasmonic excitation in every CNT in the bundle
and its characteristics would
strongly depend on  interactions between the CNTs. Two
types of such interactions have been
discussed in the literature:

(i) The first type of interaction arises from the direct coupling
of the electronic states in adjacent CNTs due to overlap of
their electron wavefunctions \cite{Maarouf}. The overlap is
always very small for two reasons: (a)  the
contact area between two adjacent CNTs is small due to their geometric  curvature,
even when the two touch each other; and (b)
the orbitals of the carbon atoms strongly overlap only in the
plane of a graphene sheet, which leads to the known van der Waals
form of the integraphene sheet interaction. The momentum mismatch
between the Fermi points of neighboring CNTs suppresses the
inter-CNT tunneling and leads to strong localization of
electronic eigenstates on individual CNTs. Therefore, there is
no significant change of low-energy band structure in the vicinity
of the Fermi energy of a compositionally disordered metallic CNT bundle
\cite{Maarouf} as compared with band structure of a single CNT.

(ii) The second type of interaction is an electrodynamic coupling in
which Coulomb interactions in a CNT are modified by the dielectric
screening induced by the adjacent CNTs. Such dielectric
screening has a significant effect in 1D structures (such as
single-wall CNTs) since many-body interactions are intrinsically
strong in 1D geometry because much of the electric field of a charge
on a CNT extends
outside of the CNT.

A bundle of parallel CNTs is generally considered as a 2D array of
infinitely long CNTs \cite{Kempa,Gumbs,Shuy}, thereby allowing the
determination of the dispersion properties of a low-frequency
plasmon in a 2D periodic medium. A realistic bundle of almost circular
cross-section has a cross-sectional diameter
much less than both its length and the electromagnetic wavelength
in free space. In order to determine the scattering and radiation
properties of a realistic CNT bundle, one needs to consider both
the  finite diameter and the length of the bundle.

Our aim in this paper is to analyze guided-wave propagation
in an almost circular bundle of metallic CNTs all of
which are either infinitely long or have finite length.
We neglect modifications of the
low-energy band structure of CNTs, but not the
electromagnetic coupling of CNTs in a bundle. Following Selpyan \emph{et al.}
\cite{Slepyan99}, we assume that, in the low-frequency
regime below optical interband transitions, the conductivity of
metalic CNTs is described by the  electron-gas model, thereby leading to the
CNT conductivity to follow the Drude model.

The rest of this paper is organized as follows. In Secs.
\ref{guided} and \ref{Effect}, a many-body technique and  an equivalent-multishell
approach, respectively, are applied to a bundle of infinitely
long metallic CNTs to derive
dispersion equations for guided-wave propagation on the bundle.
 In Sec. \ref{Finite}, scattering theory is applied
to a bundle of finite-length CNTs. Sec. \ref{num} contains
numerical results for guided-wave parameters obtained from
different approaches.

\section{Guided waves in a bundle of infinitely long CNTs}
\label{guided}

\subsection{Many-body technique} \label{problem}

Let us examine the propagation of a guided wave in an
isolated  bundle of $N$ metallic CNTs that are
closely packed together and are of infinite length.
The surrounding medium is free space (i.e., vacuum).
The effective cross-sectional diameter of the bundle is much smaller than
the wavelength in free space, and the
transverse current in all CNTs in
the bundle is neglected. An $\exp(-i\omega t)$ time-dependence
is implicit, with $i=\sqrt{-1}$, $t$ denotes the time, and $\omega$ is
the angular frequency.

 In the
low-frequency regime, only intraband transitions of
$\pi$-electrons with unchanged transverse quasi-momentum are
allowed \cite{Dresselhaus}. These transitions contribute to
the axial conductivity, but not to the transverse conductivity \cite{jnp2007},
and excite azimuthally symmetric electric
current densities in the CNTs \cite{Maksim00, Slepyan99}. At frequencies far away from
interband resonances, in practical terms, azimuthally nonsymmetric electric current densities
are not excited in CNTs because the relevant
conductivities vanish.  In detail, this peculiarity has been discussed
elsewhere \cite{Longe, Lin96} with application to surface-plasmon
modes. In the remainder of this paper, therefore we restrict ourselves to
 azimuthally symmetric electric current densities in the CNTs forming the bundle.

The superposition of the fields, induced by the electric current densities
on the surfaces of all CNTs together form a guided wave in the CNT
bundle. The electric Hertz vector ${\rm {\bf \Pi }}_m({\rm {\bf
r}}) \equiv \Pi_m({\bf r}) {\rm {\bf e}}_z $ created by the axial
electric current density on the surface of the m$^{th}$ CNT ($m\in[1,N]$) in
the bundle is governed by the Helmholtz equation
\begin{equation}
\label{eq1}
    (\nabla ^2 + k^2){\bf \Pi }_m({\bf r}) =
    {\bf 0}\,,
\end{equation}
where ${\rm{\bf e}}_z $ is the unit vector along the CNT axis (and
therefore the axis of the bundle), $k = \omega / c$,   and $c$ is
speed of light in vacuum. Let the origin of the cylindrical
coordinate system $(\rho^{(m)} ,\phi^{(m)} ,z)$ be located at the
point $z=0$ on the axis of the m$^{th}$ CNT. Since $ \Pi_m$ is a
function only of $\rho^{(m)}$ and $z$  for an azimuthally
symmetric electric current density, the nonzero components of the
electromagnetic field in this cylindrical coordinate system  are
as follows:

\begin{eqnarray}
\label{elfield1}&&    E_{\rho^{(m)}}^m = \frac{\partial ^2\Pi _m}{\partial \rho^{(m)}\partial z}\,,\\
\label{elfield} &&   E_z^m =\Bigl( {\frac{\partial ^2}{\partial z^2} + k^2} \Bigr)\Pi_m \,,\\
\label{elfield2}&&    H_{\phi^{(m)}}^m = ik\frac{\partial \Pi_m }{\partial
\rho^{(m)} }\,.
\end{eqnarray}

The electric Hertz potential $\Pi_m$  must satisfy the effective boundary
conditions \cite{Slepyan99}
\begin{eqnarray}
&\displaystyle \left.{\frac{\partial \Pi_m }{\partial \rho^{(m)}}}
    \right|_{\rho^{(m)}= R_{m} +0} -\left. {\frac{\partial \Pi_m }{\partial \rho^{(m)}}}
    \right|_{\rho^{(m)} = R_m - 0} = \nonumber
    \\
&    \rule{0in}{4ex}  \displaystyle \frac{4\pi \sigma _{m} }{ikc}
    \left[ \frac{\partial^2\Pi_m }{\partial z^2} +
    k^2\Pi_m + E^{m0}_{z}+ \sum\limits_{s = 1, s\ne m}^N {E_z^{ms}(z)}\right]\,,
         \label{boundary1}
        \\ \rule{0in}{4ex}
    & \Bigl.\Pi_m \Bigr|_{\rho^{(m)} = R_m + 0} =  \Pi_m\Bigr| _{\rho^{(m)} = R_m -
    0}\,, \label{boundary2}
\end{eqnarray}
where $\sigma _m $ is the axial conductivity of the m$^{th}$ CNT in isolation. The
scalar fields
$E^{m0}_{z}(z)$ and $E_z^{ms} (z)$ are the $z$-directed components of
the electric field on the surface of the m$^{th}$ CNT, produced by
the externally impressed sources and the axial current density on the s$^{th}$  CNT,
respectively.

Below the frequency regime of optical transitions,
an expression for the axial conductivity $\sigma_m$
is available via quantum transport theory as \cite{Slepyan99}
\begin{eqnarray}
\label{eq4} \sigma _{m} (\omega )= - \frac{ie^2}{\pi ^2\hbar R_m}
    \frac{1}{(\omega + i/\tau)}\sum\limits_{n = 1~}^{\tilde{m}}{\int\limits_{1stBZ}\!\!
    { \frac{\partial F_c}{\partial p_z}
    \frac{\partial {\cal E}_c}{\partial p_z }} }dp_z\,,
\end{eqnarray}
where $e$ is the electron charge, $\hbar $ is the normalized
Planck constant, and $p_z $ is the axial projection of the electron
quasi--momentum. The integer $n \in[1,\tilde{m}]$ labels the
$\pi $-electron energy bands, with $\tilde{m}$
appearing in the dual index ($\tilde{m},\tilde{n})$ used to
classify CNTs \cite{Dresselhaus}. The time constant $\tau$ of the
electron mean-free-path  is assumed to be equal to the
inverse relaxation frequency. The abbreviation $1stBZ$ restricts
the variable $p_z $ to the first Brillouin zone. The equilibrium Fermi distribution function
\begin{eqnarray}
\label{Fermi}
    F_{c}(p_z ,n)=
    \frac{1}{1+\exp\left[\displaystyle{
    \frac{{\cal E}_{c}(p_z,n)} {k_B^{} T}} \right] }
\end{eqnarray}
\noindent involves the temperature  $T$
and the Boltzmann constant $k_B $. The electron
energy ${\cal E}_c (p_z ,n)$ for zigzag ($\tilde{m},0$) CNTs
is   \cite{Dresselhaus}
\begin{equation}
\label{dispersive-law} {\cal E}_{c} = \gamma _0 \,\sqrt {1 + 4\cos
\left( {ap_z } \right)\cos \left( {\frac{\pi n}{\tilde{m}}}
\right) + 4\cos ^2\left( {\frac{\pi n}{\tilde{m}}} \right)}\, ;
\end{equation}
\noindent where $\gamma _0 \approx 2.7$ eV is the overlapping
integral \cite{Dresselhaus}, $a = 3b / 2\hbar$, and $b=0.142$~nm
is the interatomic distance in graphene. An expression for ${\cal
E}_c(p_z ,n)$ for armchair ($\tilde{m},\tilde{m}$) CNTs is also
available \cite{Dresselhaus}.

\subsection{Dispersion equation} \label{dispers}

A solution of Eq. (\ref{eq1}) that satisfies the
boundary condition (\ref{boundary2})
is as follows:
\begin{eqnarray}
\nonumber
&&\quad\Pi _m (\rho^{(m)},z ) =\\
&& A_m e^{ihz}
\left\{
{\begin{array}{l}
 K_0 (\kappa R_m )I_0 (\kappa \rho^{(m)} ),\mbox{ }\rho^{(m)} < R_m \\[4pt]
 I_0 (\kappa R_m )K_0 (\kappa \rho^{(m)} ),\mbox{ }\rho^{(m)} > R_m \\
 \end{array}} \right..
 \label{eq7}
\end{eqnarray}
Here $A_m$ is an amplitude, $h$ is the guide wavenumber to be
determined, and $\kappa = \sqrt {h^2 - k^2} $, while $I_0 ( \cdot )$
and $K_0 ( \cdot )$ are the modified Bessel functions of order
$0$. Equation (\ref{boundary1}) still has to be
satisfied.

\begin{figure}[!htb]
\begin{center}
\includegraphics[width=3.4in]{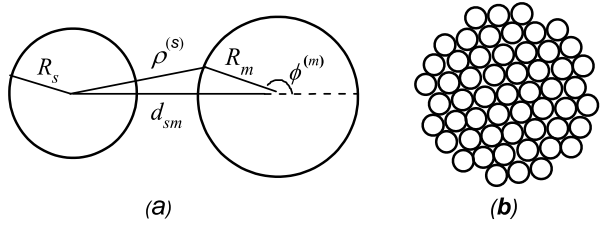}
\end{center}
\caption{(a) Schematic of two CNTs in a bundle to identify various
quantities appearing in Eq. (\ref{eq8}). (b) A bundle of $N = 55$ closely packed
CNTs that can be approximately taken to have azimuthal symmetry, and thus
has an almost circular cross-section.}
\label{fig1}
\end{figure}

The field
$E_z^{ms} (z)$ in Eq. (\ref{boundary1}) may be found by applying Eqs.
(\ref{elfield}) and ({\ref{eq7}})
in the coordinate system  $(\rho^{(s)} ,\phi^{(s)} ,z)$ to the s$^{th}$ CNT,
 $s\ne m$, and then
using the addition theorem for $K_0(\kappa \rho^{(s)})$,
$\rho^{(s)}>R_s$, to translate to the coordinate system
  $(\rho^{(m)} ,\phi^{(m)} ,z)$ \cite{Abramovitz}. With $\rho^{(s)}$ lying
  on the surface of the m$^{th}$ CNT, as shown
  in Fig. \ref{fig1}(a), we get
\begin{eqnarray}
\label{eq8} \nonumber E_z^{ms} (\rho^{(s)},z) = -A_s \kappa^2
e^{ihz}I_0 (\kappa R_s )K_0
(\kappa \rho^{(s)})= \\
 -A_s \kappa^2 e^{ihz}I_0 (\kappa R_s )\sum\limits_{\ell = - \infty
}^\infty {K_\ell (\kappa d_{sm} )I_\ell (\kappa R_m )} e^{i\ell\phi^{(m)}}\,,
\end{eqnarray}
where $d_{sm} $ is the distance between the axes of the two CNTs
and the angle $\phi^{(m)}$ has been identified in Fig.
\ref{fig1}(a).

As the conductivities for azimuthally nonsymmetric modes ($\ell
\neq 0$) are assumed to be null-valued, only the $\ell = 0$ term
in  Eqs. (\ref{eq8}) survives after the substitution that equation
into Eq. (\ref{boundary1}).
Thereafter, the substitution of Eqs. (\ref{eq7}) and
(\ref{eq8}) into Eq. (\ref{boundary1}) with $E^{m0}_{z}(z)=0$ and the subsequent
use of the Wronskian of modified Bessel functions
\cite{Abramovitz} lead to the following set of linear homogeneous equations
with unknown $A_s$, $s\in[1,N]$, written in the matrix form as
\begin{equation}
\label{eq12b}\textsf{MA} = 0\,.
\end{equation}
Here, $\textsf{A}$ is a column vector containing the $N$ unknowns
$A_s$, and the element $M_{sm}$ of the $N\times N$ matrix $\textsf{M}$
is given by
\begin{eqnarray}
\label{eq12a} \nonumber
\begin{array}{ll}
  {\displaystyle  M_{sm} =  \left\{ {\begin{array}{l}
  K_0(\kappa d_{sm})I_0 (\kappa R_m
), \quad s \ne m,\\
 K_0 (\kappa R_s ) - i\omega / [4\pi
R_s\sigma _s \kappa ^2 I_0 (\kappa R_s )], \quad s = m.
 \end{array}} \right.}
\end{array}
 \end{eqnarray}
The set  (\ref{eq12b}) of linear equations has nontrivial solutions
provided
\begin{equation}
\label{eq10} \det \textsf{M} = 0\,.
\end{equation}

The dispersion equation (\ref{eq10}) has $N$ roots
corresponding to $N$ guided waves in the CNT bundle. The
solution of
Eq. \ref{eq10} allows us to obtain the slow-wave coefficients $\beta =
k/h$ of guided waves in the CNT bundle.

\section{Equivalent-multishell approach}
\label{Effect}

The many-body technique is cumbersome for a CNT bundle with large
$N$. Furthermore, we are interested in guided waves with
azimuthal symmetry, corresponding to the low-$h$ roots of Eq.
(\ref{eq10}), as such waves are easily excited in bundles by
uniform external fields, and are also relevant for bundles of
finite-length CNTs in the long-wavelength regime (Sec.
\ref{Finite}).

An approximate but simpler method for low-$h$
guided waves can be devised for an almost circular bundle of closely packed and
\emph{identical} CNTs  as follows. Let $R_b$ stand for an effective
cross-sectional radius of the
bundle. Since the
electromagnetic field of the guided wave is azimuthally symmetric
both inside and outside the bundle, we can model the bundle as a
multishell structure comprising $\tilde{N}$ concentric shells. Each shell
in this equivalent-multishell structure
is infinitesimally thin. Thus, the
p$^{th}$ shell, $p\in[1,\tilde{N}]$, has a  cross-sectional radius
$\tilde{R}_p $  such that
$R_b = \tilde{R}_{\tilde{N}} > \tilde{R}_{\tilde{N} - 1}
> ... > \tilde{R}_1$. The effective surface conductivity of the p$^{th}$ shell is given by $\tilde {\sigma }_p  =
{\Sigma}_p / (2\pi \tilde{R}_p )$, where ${\Sigma}_p $ is equal to
the sum of the axial conductivities of all CNTs placed between the
p$^{th}$ and the (p$-1$)$^{th}$ shells. There is some latitude
inherent in the procedure to select $\tilde{N}$, $\tilde{R}_p$,
and $\tilde{\sigma}_p$, $p\in[1,\tilde{N}]$.

Let $\Pi_p(\rho,\phi,z)$ be the Hertz potential everywhere \emph{entirely} due to the
p$^{th}$ shell, with the cylindrical coordinate system $(\rho,\phi,z)$
located at the point $z=0$ on the axis of the CNT bundle.
The boundary conditions across  the p$^{th}$ shell due to
the axial current density ${\bf J}_p(z)=J_p(z) {\bf e}_z$ on the surface $\rho =\tilde{R}_p$ are
as follows:
\begin{eqnarray}
&\label{boundary3} \displaystyle \left.{\frac{\partial \Pi_p
}{\partial \rho}}
    \right|_{\rho= \tilde{R}_{p} +0} -\left. {\frac{\partial \Pi_p }{\partial \rho}}
    \right|_{\rho = \tilde{R}_p - 0} = \frac{4\pi}{ikc} J_p\,,
    \\ \rule{0in}{4ex}
    & \Bigl.\Pi_p \Bigr|_{\rho = \tilde{R}_p + 0} =  \Pi_p\Bigr| _{\rho = \tilde{R}_p -
    0}\,. \label{boundary4}
\end{eqnarray}
Here
\begin{equation} \label{current}
J_p(\phi,z) = \tilde{\sigma}_p
    \left[ \sum\limits_{q = 1}^{\tilde{N}} {\left[ \frac{\partial^2}{\partial z^2} +
    k^2\right] \Pi_q (\tilde{R}_p, \phi,z)}  + E^0_{z} (z) \right]\,,
\end{equation}
where $E^0_z$ is the $z$-directed component of the externally impressed
electric field.

With $A_p^{(\ell)}$, $\ell\in[0,\infty)$, representing its amplitude,
an expression for $\Pi_p$  is as follows:
\begin{equation}
\label{eq13} \Pi _p (\rho ,\phi,z ) = A_p^{(\ell)} e^{ihz}e^{i\ell\phi
}\left\{ {\begin{array}{l}
 K_\ell (\kappa \tilde{R}_p )I_\ell (\kappa \rho ),\mbox{ }\rho < \tilde{R}_p , \\
 I_\ell (\kappa \tilde{R}_p )K_\ell (\kappa \rho ),\mbox{ }\rho > \tilde{R}_p . \\
 \end{array}} \right.
\end{equation}
Substitution of
(\ref{current}) with $E^0_z=0$ and  Eqs. (\ref{eq13})  into Eq. (\ref{boundary3}) leads to
a set of linear homogeneous equations with unknown
$A_p^{(\ell)}$, $p\in[1,\tilde{N}]$. This set has nontrivial solutions that can be
ascertained by solving
dispersion equation
\begin{equation}
\label{eq12} \det \tilde{\textsf{M}} = 0\,
\end{equation}
 for the determination of $h$.
The element $\tilde{M}_{qp}$ of the
$\tilde{N}\times\tilde{N}$ matrix $\tilde{\textsf{M}}$ is to be computed as
\begin{eqnarray}
 \nonumber
\begin{array}{ll}
  {\displaystyle  \tilde{M}_{qp} =  \left\{ {\begin{array}{l}
 K_\ell (\kappa \tilde{R}_q )I_\ell (\kappa \tilde{R}_p ), \quad q < p, \\
 K_\ell (\kappa \tilde{R}_p )I_\ell (\kappa \tilde{R}_q ), \quad q > p,\\
 K_\ell (\kappa \tilde{R}_q )I_\ell
 (\kappa \tilde{R}_q ) - i\omega / [4\pi \tilde{R}_q \tilde {\sigma }_q \kappa ^2], \quad q = p. \\
 \end{array}} \right.}
\end{array}
 \end{eqnarray}
For almost circular bundles with $N=55$ CNTs,
we
compared the first three solutions of Eq. (\ref{eq10}) with those
of Eq. (\ref{eq12}) for $\ell=0$, and obtained good agreement, as
discussed in Sec. \ref{num}.

In contrast to an isolated CNT and even a bundle with relatively
small number of closely packed CNTs, in the low-frequency regime
an azimuthally nonsymmetric guided wave with $\ell \ne 0$ can
exist in a CNT bundle with a large  number of CNTs.
 This guided wave is formed by the ensemble of azimuthally symmetric
electric current densities excited in every CNT of the bundle. The
local field of the azimuthally nonsymmetric wave quickly changes
in the central part of the bundle. Since symmetric electric
current densities are excited by the spatially homogeneous
component of the local field, then CNTs in the central core of the
bundle practically are not excited and their conductivity may be
supposed to be zero. Therefore, in order to approximately describe
an azimuthally asymmetric guided wave by Eqs. (\ref{eq13}) and
(\ref{eq12}), the conductivity of the inner shells with radius
$\tilde{R}_p < 10\ell R_0 / \pi $ should be assumed to be equal to
zero, where $R_0$ is the radius of every CNT in the bundle. It is
expected that an azimuthally nonsymmetric guided wave can be
easily excited in a bundle of parallel closely placed CNTs and
contribute greatly to interbundle interactions.

\section{Guided waves in a bundle of finite-length CNTs}
 \label{Finite}

In order to investigate the finite-length effects in CNT bundles, let us apply
 integral-equation methods  developed for a single CNT
\cite{Slepyan06, Hanson05} and for a planar array of CNTs
\cite{Hanson06, Hanson07}. Let a closely packed bundle of parallel
and identical CNTs of length $L$ be aligned parallel to the $z$
axis of a Cartesian coordinate system. The bundle has an almost
circular cross-section so the bundle radius $R_b $ can be
prescribed; $R_b$ is assumed here to be much less than the
wavelength $\lambda $ and length $L$. On exposure to an externally
impressed field that is almost homogeneous over the bundle
cross-section, an axial and azimuthally symmetric surface current
density is excited in every CNT. Following   Sec. \ref{Effect}, we
replace the CNT bundle by $\tilde{N}$ multishells, and prescribe
the radius $\tilde{R}_p$ as well as the effective axial
conductivity $\tilde{\sigma}_p$ of the p$^{th}$ shell,
$p\in[1,\tilde{N}]$. The surface current density, induced by the
externally impressed field on the surface of the p$^{th}$ shell,
is denoted by ${\rm {\bf J}}_p (z) = J_p (z){\rm {\bf e}}_z $. It
has to satisfy the edge conditions
\begin{equation}
\label{eq14} J_p (\pm L/2) = 0,
\end{equation}
which express the
absence of concentrated charges on the two edges of the bundle.

The electric Hertz potential  $\Pi (\rho,z )$
satisfies the Helmholz equation (\ref{eq1}),
the radiation condition
\cite{Weinstein}, as well as the
the
boundary conditions (\ref{boundary3}) and (\ref{boundary4});
the cylindrical coordinate system $(\rho,\phi,z)$
located at the point $z=0$ on the axis of the CNT bundle.
The potential $\Pi (\rho,z )$ is expressed in the form of  a single-layer potential as
\begin{equation}
\label{eq15} \Pi (\rho,z ) = \frac{i}{\omega }\sum\limits_{p =
1}^{\tilde{N}} {\tilde{R}_p \int\limits_{ - L / 2}^{L / 2} {J_p
({z}')G (z - {z}',\rho ,\tilde{R}_p )d{z}'} } \,,
\end{equation}
where
\begin{equation}
\label{eq16} G(z,\rho ,R) = \int\limits_{0}^{2\pi} {\frac{\exp
\left\{ {ik\sqrt {\rho ^2 + R^2 - 2R\rho \cos \varphi + z^2} }
\right\}}{\sqrt {\rho ^2 + R^2 - 2R\rho \cos \varphi + z^2}
}d\varphi }
\end{equation}
\noindent is the free-space scalar Green function and
\begin{equation}
\label{eq17} J_ p(z) = \tilde{\sigma}_p \left[ \frac{\partial
^2\Pi ( \tilde{R}_p,z )}{\partial z^2} + k^2\Pi (\tilde{R}_p,z ) +
E^0_{z}(z) \right].
\end{equation}

Setting $\rho = \tilde{R}_p $ in Eq. (\ref{eq15}) and making use of
Eq. (\ref{eq17}), we obtain a system of $\tilde{N}$
integral equations with respect to the unknown current densities
as follows:
\begin{eqnarray}
\nonumber &&
\Phi _p (z) = \sum\limits_{q = 1}^{\tilde{N}} {\int\limits_{ - L /
2}^{L / 2} { \left\{ {\frac{i2\pi \tilde{R}_q }{\omega }G(z -
{z}',\tilde{R}_q
,\tilde{R}_p )} \right.} }\\[5pt]
\nonumber
 && \qquad -
 \left. {\frac{\delta _{qp} }{2ik \tilde{\sigma} _q }\exp (ik\vert
z - {z}'\vert )} \right\}\\[5pt]
\label{eq18}
&&\qquad\qquad\times\,J_q ({z}')d{z}'\,,\quad p\in[1,\tilde{N}]\,.
\end{eqnarray}
Here,
\begin{eqnarray}
\label{eq19} \nonumber &&
 {\displaystyle  \Phi _p (z) = - \frac{1}{2ik\tilde{\sigma}_p }\int\limits_{
- L / 2}^{L / 2} {E^0_{z} ({z}')\exp (ik\vert z - {z}'\vert
)d{z}'} }\\[5pt]
 &&
 {\displaystyle\quad + C_p \exp (ikz) + D_p \exp ( - ikz)}\,
 \end{eqnarray}
 and
$\delta _{qp} $ is the Kronecker delta, whereas $C_p $ and $D_p $
are unknown constants to be determined from the edge conditions
(\ref{eq14}). Parenthetically, the system (\ref{eq18}) with a
different Green function was applied to a planar array of finite
CNTs \cite{Hanson06, Hanson07}.

The integral on the right side of
(\ref{eq18}) can be numerically handled by a quadrature formula, thereby
transforming the system (\ref{eq18}) into a matrix equation. The solution of
the corresponding characteristic equation yields eigenfrequencies
and eigenmodes of a finite coaxial cylinder as a high-Q
microcavity.

In the long-wavelength regime ($\lambda \gg L )$, the
electromagnetic properties of the CNT bundle can be characterized by
the polarizability scalar
\begin{equation}
\label{eq20} \alpha = \frac{2\pi i}{\omega E^0_{z}(0)
}\sum\limits_{p = 1}^{\tilde{N}} {\tilde{R}_p \int_{ - L / 2}^{L /
2} {J_p (z)dz} }
\end{equation}
As shown elsewhere \cite{Slepyan06, Hanson05, Burke06},
an isolated CNT can function as an antenna in the terahertz regime wherein the
CNT has geometrical resonances of guided wave with slow-wave
coefficient $\beta_0$ at frequencies related to the CNT length $L$
by the condition
\begin{equation}
\label{eq20a} Lk \approx \pi \tilde{s}\, {\rm Re}(\beta_0), \quad
\tilde{s} = 1,2,\dots\,,
\end{equation}
where $\beta_0$ is the slow-wave coefficient for an isolated CNT.
The antenna effect of an array of multiwall CNTs was
experimentally found at a frequency satisfying the condition
(\ref{eq20a}) with $\tilde{s}=1$ and $\beta_0 \approx 1$
\cite{Wang, Kempa07}.

In this paper, we are interested in the antenna efficiency of a
CNT bundle at the first antenna resonance: $\tilde{s}=1$ in Eq.
(\ref{eq20a}). The antenna efficiency is defined as the ratio
\begin{equation}
\label{eq21} \eta = \frac{P_r }{P_t + P_r }\,,
\end{equation}
\noindent where
\begin{equation}
\label{eq21a} P_r = \frac{\pi^2 \omega^2 }{c^3}{\int_0^\pi
\sin^3\theta
 \left|
  { \int_{ -
L / 2}^{L / 2}  { e^{ikz\cos \theta }} {\sum\limits_{p =
1}^{\tilde{N} } \tilde{R}_p J_p (z) dz } } \right|^2 d\theta}
\end{equation}
 is the radiated
power and
\begin{equation}
\label{eq21b} P_t = \pi {\rm Re} \left( {\sum\limits_{p = 1}^{\tilde{N} }
{\frac{\tilde{R}_p}{\tilde{\sigma _p} }\int_{ - L / 2}^{L / 2}
{\vert J_p (z)\vert ^2dz} } } \right)
\end{equation}
is the power lost to ohmic dissipation.

In the long-wavelength regime, the antenna efficiency $\eta _0 $
of an isolated CNT is equivalent to that of a dipole antenna of
length $L$ and resistance per unit length $R_{dip} $
\cite{Burke06}; i.e.,
\begin{equation}
\label{eq22} \eta _0 = \frac{L / \lambda }{L / \lambda + 3cR_{dip}
\lambda / (8\pi^2 )}\,,
\end{equation}
where $R_{dip}= {\rm Re}[1/(2\pi R_0 \sigma_0)]$ for an isolated
CNT of cross-sectional radius $R_0$ and surface conductivity
$\sigma_0$. The high value of $R_{dip} $ \cite{Burke06-2} and the
small value of $L / \lambda $ at the first antenna resonance lead
to a very small antenna efficiency of an isolated single CNT,
i.e., $\eta_0 \approx 10^{ - 4} - 10^{ - 6}$
\cite{Burke06,Hanson05}.

However, the situation is more optimistic for a bundle of $N$
metallic CNTs. Such a CNT bundle can be considered as a composite
antenna containing $N$ in-phase parallel dipole antennas, so that
$P_r \sim N^2$ but  $P_t \sim N$ from Eqs. (\ref{eq21a}) and
(\ref{eq21b}). Then the antenna efficiency of the CNT bundle at
the frequency of first geometrical resonance of the axially
symmetric guided wave with the highest slow-wave coefficient
$\beta$ is
\begin{equation}
\label{eq23} \eta \approx \frac{N \eta _0 \,{\rm Re}(\beta) }{\eta
_0 [N \,{\rm Re} (\beta) - {\rm Re} (\beta_0)] + {\rm Re} (\beta_0) },
\end{equation}
where $\eta _0$ is the antenna efficiency of an isolated CNT at
the first antenna resonance at the same frequency.

Because of electromagnetic coupling of the CNTs in a bundle with
$N\gg1$ CNTs, the inequality ${\rm Re}(\beta )\gg {\rm
Re}(\beta_0)$ holds true, as shown in Sec. \ref{num}; therefore,
$\eta \gg \eta_0$.

\section{Characteristics of guided waves} \label{num}

Calculations were performed for almost circular  bundles made of $N$ parallel,
identical,  single-wall, metallic, zigzag (21,0) CNTs. The
relaxation time was taken as $\tau = 10^{-13}$ s. For convenience, the CNTs were assumed to be
close-packed on a 2D triangular lattice with intertube spacing 3.4 \AA~ as
shown in Fig. \ref{fig1}(b) for $N=55$.

\subsection{Guided waves in bundles of infinitely long CNTs}
\label{num1}

\begin{figure}[!htb]
\begin{center}
\includegraphics[width=3.4in]{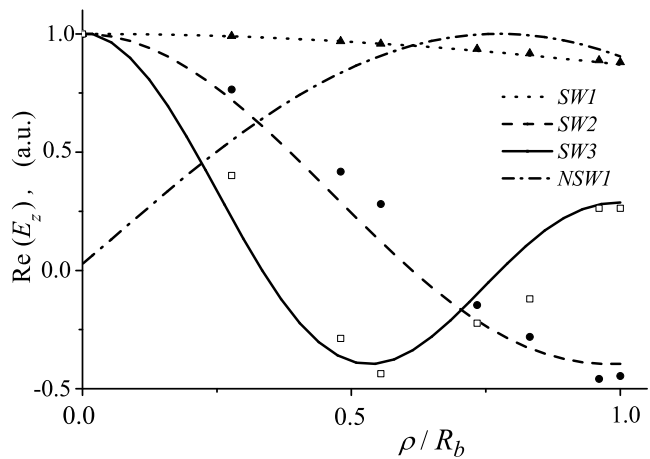}
\end{center}
\caption{Radial dependencies of the magnitude of the $z$-directed
component of the electric field (in arbitrary units) of the guided
waves in a CNT bundle of $N=55$ parallel, identical, infinitely
long, single-wall, metallic, zigzag (21,0) CNTs arranged on a
triangular lattice. The points correspond to the solutions of Eq.
(\ref{eq10}), and the lines to the solutions of Eq. (\ref{eq12}).}
\label{fig2}
\end{figure}

Let us begin with numerical results for bundles of infinitely long CNTs.
  We considered only three
roots each of the dispersion equations (\ref{eq10}) and (\ref{eq12}). These roots~---~labeled
$h_1
$, $h_2 $ and $h_3 $~---~were the ones with the smallest real parts
(${\rm Re}(h_3 ) > {\rm Re}(h_2 )
> {\rm Re}(h_1 ))$, and correspond to azimuthally  symmetric
guided waves  identified as $SW1$, $SW2$ and $SW3$, respectively.
Only these guided waves mostly influence the scattering
properties of finite-length CNT bundles, as discussed in Sec.
\ref{num2}. We also considered the properties of an azimuthally
nonsymmetric guided wave, identified as $NSW1$, which emerges
from the solution of Eq. (\ref{eq12}) for $\ell = 1$ and the real
part of whose wavenumber $h$ is the smallest possible.

The radial dependencies of the magnitude of the $z$-directed
component of the electric field inside the chosen CNT bundle for
guided waves $SW1$, $SW2$, $SW3$ and $NSW1$ are shown in Fig.
\ref{fig2}. The lines in this figure were obtained from the
effective multishell approach of Sec. \ref{Effect}, whereas the
points were obtained from the many-body technique of Sec.
\ref{dispers}.  We conclude that the two approaches yielded
reasonably close results for the chosen bundle. Such a good
coincidence of the results from both approaches was observed only
for bundles with closely packed CNTs. For rarified bundles (where
the smallest inter-CNT distance exceeds the CNT radius by a factor
of 10 or more), the results of both approaches coincide only for
the guided wave $SW1$.

\begin{figure}[!htb]
\begin{center}
\includegraphics[width=3.4in]{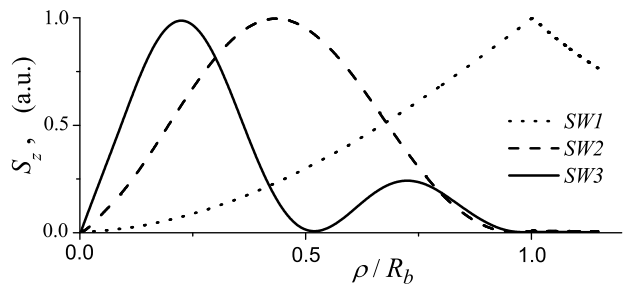}
\end{center}
\caption{Radial dependence of the axial component of the
time-averaged Pointing vector $S_z$ (in arbitrary units) of the
guided waves in the same CNT bundle as in Fig. \ref{fig2}.}
\label{fig3}
\end{figure}

\begin{figure}[!htb]
\begin{center}
\includegraphics[width=3.4in]{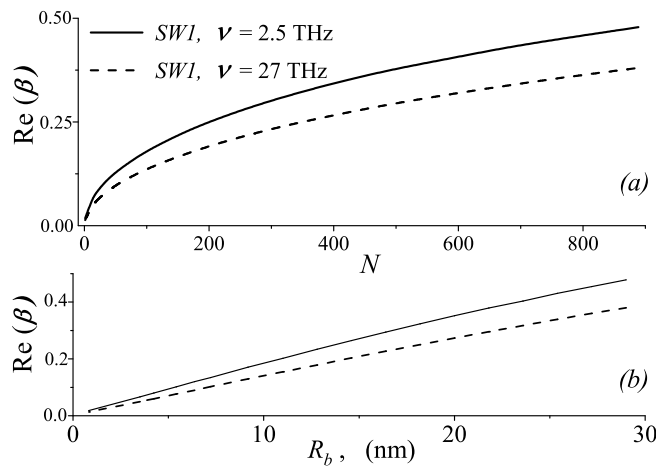}
\end{center}
\caption{Dependencies of ${\rm Re}(\beta)$ of the guided wave SW1
on (a) the number $N$ of CNTs in the bundle  and (b) the bundle
radius $R_b$.} \label{fig4}
\end{figure}

As one can be deduced from Fig. \ref{fig2}, the electric field
inside the bundle is distributed over the entire cross--section.
Outside the bundle, the radial distribution of the electric field
is governed by the decreasing modified Bessel function
$K_\ell(\sqrt {h^2 - k^2} \rho)$.  Thus the electric field is
highly localized to the CNT bundle.

Furthermore, calculations show that the radial field distribution
for $SW2$ and $SW3$ can be described adequately by Bessel
functions of the first kind  $J_0 (\kappa _2 \rho )$ and $J_0
(\kappa _3 \rho )$, respectively, where $\kappa _2 $ and $\kappa
_3>\kappa_2 $ are the non-zero minimal roots of the equation
 $J_1 (\kappa R_b ) = 0$.

\begin{figure}[!htb]
\begin{center}
\includegraphics[width=3.4in]{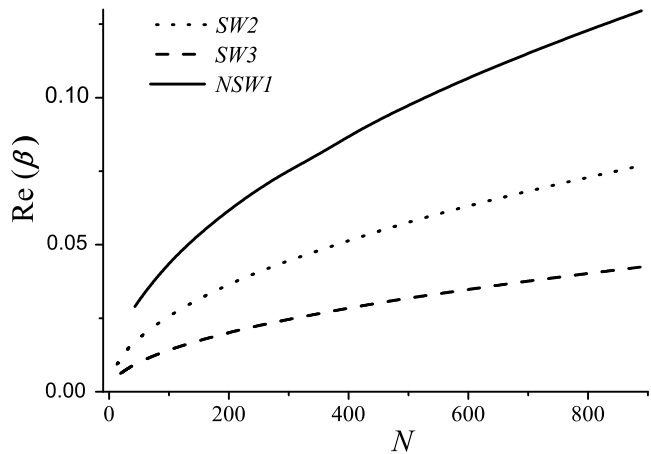}
\end{center}
\caption{Dependence of ${\rm Re}(\beta)$ on $N$, for the guided
waves SW2, SW2 and NSW1 at $\nu= \omega/2\pi= 27$~THz.}
\label{fig5}
\end{figure}

\begin{figure}[!htb]
\begin{center}
\includegraphics[width=3.4in]{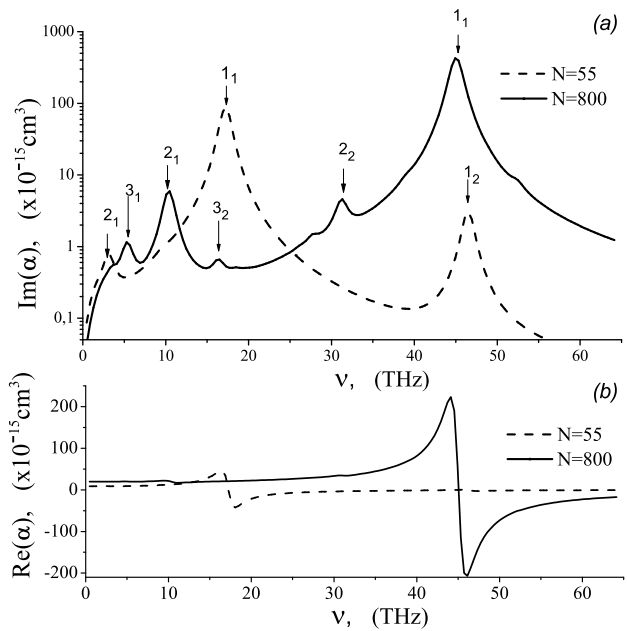}
\end{center}
\caption{Frequency dependence of (a) ${\rm Im}(\alpha)$ and (b)
${\rm Re}(\alpha)$ of a bundle of CNTs of length $L=500$~nm. Two
values of $N$ were used: 55 and 800. The labels 1, 2 and 3 denote
the geometric resonance of the guided waves $SW1$, $SW2$ and
$SW3$, respectively. The subscripts on the labels denote the
number $\tilde{s}$ of the geometric resonance in Eq.
(\ref{eq20a}).} \label{fig6}
\end{figure}

All other results~---~presented in Fig.
\ref{fig3}-\ref{fig7}~---~were obtained using the effective
multishell approach.  Figure~\ref{fig3} shows the radial
distributions of the axial component of the time-averaged Pointing
vector $S_z=c|H_\phi|^2/(8\pi)$ for three azimuthally symmetric
guided waves ($SW1$, $SW2$, and $SW3$) inside  the CNT bundle
($\rho/R_b<1$) and in the vicinity of its surface ($\rho/R_b>1$).
As shown in Fig.~\ref{fig3}, the axial component of the
time-averaged Pointing vector of $SW1$ is maximum near the surface
of the bundle, and a large part of the energy of SW1 leaks outside
the bundle. In contrast, the power densities of $SW2$ and $SW3$
are mostly concentrated inside the bundle. Thus, the
electromagnetic energy and volume electric current density of the
guided waves in a CNT bundle are distributed similarly to those of
eigenwaves propagating in a macroscopic, lossy, infinitely long
wire \cite{Sommerfeld}. Accordingly, guided waves propagating in a
CNT bundle with a large number of CNTs cannot be called
\emph{surface waves}.

Figure \ref{fig4}(a) contains plots of the real part of the
slow-wave coefficient $\beta$ of the guided wave $SW1$ in the
terahertz ($\nu=\omega/2\pi = 2.5$ THz) and infrared ($\nu = 27$
THz) regimes for different numbers of CNTs in the bundle. The
coefficient ${\rm Re}(\beta)$ increases as much as 26 times with
$N$ increasing up to 900. A large $N$ means that the bundle is
thick (with $R_b > 25$ nm) and its slow-wave coefficient tends to
that of a macroscopic metallic wire. The dependence of ${\rm
Re}(\beta)$ on the bundle radius is linear up to $R_b = 25$ nm, as
may be deduced from Fig. \ref{fig4}(b).

A comparison of Figs. \ref{fig4}(a) and Fig. \ref{fig5} reveals
that ${\rm Re}(\beta) $ of   $SW1$ is about three, five and nine
times more than that of $NSW1$, $SW2$ and $SW3$, respectively. We
also found that the solution of the dispersion equation
(\ref{eq12}) for $NSW1$ does not change in a frequency regime
wherein the arguments of Bessel functions $I_\ell (\cdot)$ and
$K_\ell (\cdot)$ in Eq. (\ref{eq13}) are much smaller than unity,
i.e. $\kappa R_b < < 1$. This condition holds for realistic CNT
bundles over a wide frequency range from the terahertz to the
near-infrared regimes \cite{Slepyan99}. The dependencies of ${\rm
Re}(\beta)$ on $R_b$ are linear for $SW2$, $SW3$ and $NSW1$ are
also linear up to $R_b = 25$ nm (not shown in this paper).

Although not substantiated here by a graph, the value of $-{\rm
Im} (\beta)/{\rm Re}( \beta)$ for the considered guided waves does
not depend on $N$.
 This implies that the
electromagnetic coupling of the CNTs does not influence the
attenuation of the guided waves in the bundle. The ratio
 $-{\rm Im} (\beta)/{\rm Re}( \beta)$
is approximately equal to 0.32 and 0.034 at
$\nu = 2.5$ THz and $\nu = 27$ THz, respectively, for $SW1$.
\subsection{Guided waves in bundles of finite-length CNTs}
 \label{num2}
 Let us now move on to almost circular bundles of finite-length CNTs.
Fig. \ref{fig6} demonstrates the frequency dependence of the
polarizability scalar $\alpha$ of a bundle of  CNTs for two
different values $N$. The labels $1$, $2$ and $3$  in this figure
denote the geometric resonance of the guided waves $SW1$, $SW2$
and $SW3$, respectively. Clearly, the polarizability resonances in
this figure occur at frequencies satisfying the condition
(\ref{eq20a}). The location of the first resonance ($\tilde{s} =
1)$ of all three guided waves on the frequency axis depends on
$N$. The first geometrical resonance of the $SW1$ is the strongest
of the three, and it shifted from the terahertz regime to the
mid-infrared regime as $N$ was changed from 1 to 800.
The geometrical resonances of $SW2$ and $SW3$ occur at lower
frequencies; moreover, they even vanish for small $N$, because of
the strong attenuation of the guided wave at low frequencies
(where the condition ${\rm Im}(h)/{\rm Re}(h) > 1$ holds).

\begin{figure}[!htb]
\begin{center}
\includegraphics[width=3.4in]{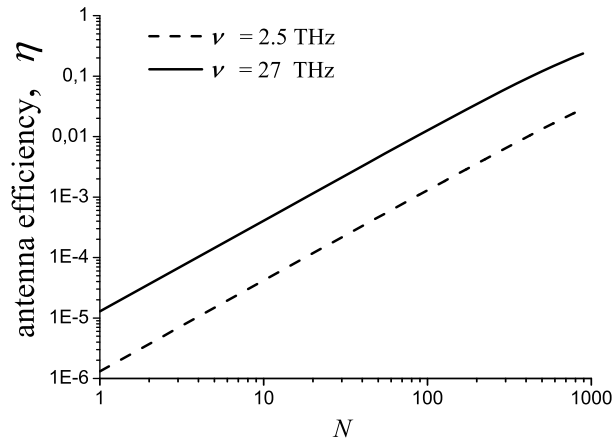}
\end{center}
\caption{Dependence of the antenna efficiency $\eta$ on the number $N$
of CNTs  at $\nu = 2.5$ THz (terahertz regime) and $\nu =
27$ THz (infrared regime). The length $L$ was modified to ensure that the condition  (\ref{eq20a}) is satisfied with $\tilde{s}=1$, for fixed $\nu$.
} \label{fig7}
\end{figure}

Similar conclusions could be made for
wavelength-dependent resonant effects in the optical regime from
measurements on a parallel array of multiwall CNTs
\cite{Wang}. Thus, by varying the bundle radius and the CNT length, one
can tune the resonance properties of a CNT bundle functioning as an antenna
 from the
terahertz to the visible regimes. A similar conclusion could be drawn for
composite materials containing CNT inclusions \cite{Slepyan06,lakhtakia}.

The dependence of the antenna efficiency $\eta$
on the number $N$ of CNTs in a bundle at two different frequencies $\nu=\omega/2\pi$
is illustrated in Fig. \ref{fig7}.
The length $L$ was modified to ensure that a chosen value of $\nu$
always corresponds to the first geometrical resonance of SW1 (i.e.,
the condition  (\ref{eq20a}) is satisfied with $\tilde{s}=1$).
Evidently from Fig. \ref{fig7}, $\eta$ increases
with $N$ and tends to unity for thick bundles ($N
> 800$); indeed, $\eta = 0.24$ when $N = 900$
and $\nu = 27$ THz.

\section{Concluding remarks}
To conclude, an equivalent-multishell approach was proposed for the
approximate calculation of the characteristics of electromagnetic guided waves
on almost circular, closely packed  bundles of parallel, identical,
and metallic carbon nanotubes. The CNTs can be either infinitely
long or of finite length.
 The
dispersion characteristics of the guided waves with the smallest
retardation (i.e., lowest values of ${\rm Re} (h)$) where studied
for bundles of infinitely long CNTs. The slow-wave coefficients
for azimunthally symmetric guided waves were found to increase
with the number of metallic CNTs in the bundle, tending  for thick
bundles to unity, which is characteristic of macroscopic metallic
wires. The existence of an azimuthally nonsymmetric guided wave
at low frequencies in a bundle of a large number of finite-length
CNTs was demonstrated, in contrast to the characteristics of
guided-wave propagation in a single CNT.

The polarizability scalar and the antenna efficiency of a bundle
of  finite-length CNT in the long-wavelength regime were
calculated over a wide frequency range spanning the terahertz and
the near-infrared regimes. The resonances of different guided
waves in  CNT bundles caused by edge effects (geometrical
resonances)  were identified. The antenna efficiency  of a CNT
bundle at the first antenna resonance can greatly exceed that of a
single CNT. Thus, the analysis carried out in this paper forms a
basis for the design and development of CNT-bundle antennas and
composite materials \cite{Lakht96} containing CNT-bundles as
inclusions.

\begin{center}{\bf ACKNOWLEDGMENTS} \end{center}

The authors are grateful to Dr. G. Ya. Slepyan for helpful
discussions. The research of MVS and SAM was partially supported
by the NATO Science for Peace program (grant SfP-981051), the
State Committee for Science and Technology of Belarus and the
INTAS (grant 03-50-4409), the INTAS (grant 05-1000008-7801) and
the Belarus Republican Foundation for Fundamental Research and
Russian Foundation for Basic Research (grant F06R-101, F07M-069).
The research of AL was partially supported by the Charles Godfrey
Binder Professorship Endowment at the Pennsylvania State
University.


\begin{thebibliography}{99}

\bibitem{WL06} {\em Selected Papers on Nanotechnology~--~Theory
and Modeling\/} (F. Wang and A. Lakhtakia, eds),  (SPIE Press,
Bellingham, WA, 2006).

\bibitem {Longe} P. Longe, and S. M. Bose, Phys. Rev. B \textbf{48}, 18239 (1993).

\bibitem {Slepyan99} G. Ya. Slepyan, S. A. Maksimenko, A. Lakhtakia, O. Yevtushenko, and A. V. Gusakov,  Phys. Rev. B \textbf{60}, 17136 (1999).

\bibitem {Slepyan06} G. Ya. Slepyan, M. V. Shuba, S. A. Maksimenko,
and A. Lakhtakia, {Phys. Rev. B} \textbf{73}, 195416 (2006).

\bibitem {Hanson05} G. W. Hanson, {IEEE Trans. Antennas Propagat.} \textbf{53}, 3426 (2005).

\bibitem {Burke06} P. J. Burke, S. Li, and Z. Yu, IEEE Trans. Nanotechnol. \textbf{5}, 314
(2006).

\bibitem {Maarouf} A. A. Maarouf, C. L. Kane, and E. J. Mele, Phys. Rev. B \textbf{61}, 11156 (2000)

\bibitem {Kempa} K. Kempa, Phys. Rev. B \textbf{66}, 195406 (2002).

\bibitem {Gumbs} G. Gumbs, and G. R. Aizin, Phys. Rev. B \textbf{65}, 195407 (2002).

\bibitem {Shuy} F. L. Shyu, and M. F. Lin, Phys. Rev. B \textbf{62}, 8508
(2000).

\bibitem {Dresselhaus} M. S. Dresselhaus, G. Dresselhaus, and Ph. Avouris, \textit{Carbon Nanotubes} (Springer,
Berlin, 2001).

\bibitem{jnp2007}
S. A. Maksimenko, A. A. Khrushchinsky, G. Ya. Slepyan, and P. V.
Kibis, J. Nanophoton. \textbf{1}, 013505 (2007).

\bibitem{Maksim00} S. A. Maksimenko and G. Ya. Slepyan,
in \textit{Electromagnetic Fields
in Unconventional Materials and Structures}, edited by O. N. Singh and
A. Lakhtakia (Wiley, New York, 2000), pp. 217-255.

\bibitem {Lin96} M. F. Lin, D. S. Chuu, C. S. Huang, Y. K. Lin, and K. W.-K. Shung, Phys. Rev. B \textbf{53}, 15493
(1996).

\bibitem {Abramovitz} M. A. Abramovitz, and I. A. Stegun, \textit{Handbook of Mathematical Functions} (Dover
Press, New York, 1972).

\bibitem {Hanson06}  J. Hao, and G. W. Hanson, {Phys. Rev. B} \textbf{74}, 035119 (2006).

\bibitem {Hanson07}  J. Hao, and G. W. Hanson, {Phys. Rev. B} \textbf{75}, 165416 (2007).

\bibitem {Weinstein} L. A. Weinstein, \textit{The Theory of Diffraction and the Factorization Method}
(Golem, New York, 1969).

\bibitem {Wang} Y. Wang, K. Kempa, B. Kimball, J. B. Carlson, G. Benham, W. Z. Li, T.
Kempa, J. Rybczynski, A. Herczynski, and Z. F. Ren, Appl.
Phys. Lett.  \textbf{85}, 2607 (2004).

\bibitem {Kempa07} K. Kempa, J. Rybczynski, Z. Huang, K. Gregorczyk, A. Vidan,
B. Kimball, J. Carlson, G. Benham, Y. Wang, A. Herczynski, and Z.
Ren, Adv. Mater.  \textbf{19}, 421 (2007).

\bibitem {Burke06-2} Z. Yu, C. Rutherglen, and P. J. Burke, Appl.
Phys. Lett.  \textbf{88}, 233115 (2006).

\bibitem {Sommerfeld} A. Sommerfeld, \textit{Electrodynamics} (Academic, New York,
1952).

\bibitem{lakhtakia} A. Lakhtakia, G. Ya. Slepyan, S. A. Maksimenko, O. M. Yevtushenko,
A. V. Gusakov,  Carbon \textbf{36}, 1833 (1998).

\bibitem{Lakht96} {\em Selected Papers on Linear
Optical Composite Materials\/} (A. Lakhtakia, ed),  (SPIE Press,
Bellingham, WA, 1996).


\end{thebibliography}
\end{document}